\documentclass[a4paper,11pt]{article}

\usepackage{authblk}

\usepackage[text={174mm,258mm},%
  papersize={210mm,297mm},%
columnsep=12pt,%
headsep=21pt,%
centering]{geometry}
\usepackage[backend=biber,style=nature,autocite=superscript,url=false,doi=true,sortcites=false,eprint=false,isbn=false]{biblatex}
\usepackage{url}

\usepackage{graphicx}%
\usepackage{multirow}%
\usepackage{amsmath,amssymb,amsfonts}%
\usepackage{amsthm}%
\usepackage{mathrsfs}%

\usepackage[title]{appendix}%
\usepackage{xcolor}%
\usepackage{textcomp}%
\usepackage{manyfoot}%
\usepackage{booktabs}%
\usepackage{algorithm}%
\usepackage{algorithmicx}%
\usepackage{algpseudocode}%
\usepackage{listings}%

\title{Disentangling Large-Scale Supply Networks: f-HiCoNE Framework for Flow-Hierarchical Clustering via Combinatorial Hodge Decomposition}

\author[1]{Taiyo NAKATANI$^*$}
\author[2]{Takaaki Aoki}

\affil[1]{Graduate School of Data Science, Shiga University, Hikone, Shiga 522-8522, Japan.}
\affil[2]{Faculty of Data Science, Shiga University, Hikone 522-8522, Japan.}

\date{Corresponding author: tnakatani2000@gmail.com}

\begin{document}
\maketitle

\begin{abstract}
    Modern society relies on complex supply chains to sustain the flow of goods and services that are essential to daily life. While traditional supply chain theory assumes a clear, hierarchical flow from upstream suppliers to downstream customers, observable real-world transaction networks rarely exhibit this acyclic structure. Instead, detailed inter-firm data reveal that interwoven networks are heavily entangled by cyclic flows. Consequently, without appropriate partitioning of these massive inter-firm networks, the latent flow-hierarchical structures that are central to supply chain concepts remain obscure. To address this analytical challenge, we introduce the flow-Hierarchical Community Network Extraction (f-HiCoNE) framework. By applying combinatorial Hodge decomposition, this approach disentangles the complex inter-firm network by isolating the acyclic gradient flow to quantify the flow-hierarchical parts and partition the graph. By applying f-HiCoNE to a nationwide transaction dataset of approximately 650,000 firms, we successfully extracted functional supply-chain clusters. These clusters demonstrated strong flow-hierarchical organisation, wherein the upstream-downstream positioning of firms was accurately captured by local scalar potentials, revealing distinct geographically localised industrial ecosystems. This study provides a map that helps firms understand their surrounding environment and locate their position within an inter-firm network and opens a new research avenue focused on flow-hierarchy clustering in supply chain analysis.
\end{abstract}


\section{Introduction}\label{sec1}
Supply chains are essential for producing goods and delivering services because few outputs are produced by a single firm end-to-end; instead, value is created and moved through many firms connected by supplier-buyer relationships.
As production becomes more distributed across organizations and locations, these relationships collectively form the backbone of everyday economic activity and shape what can be produced, where, and at what speed \cite{mena2013theory_multitier_supply,bellamy2013network_analysis_supply,diem2024estimating_loss_economic}.
In the face of growing global uncertainty and economic turbulence, a systematic understanding of supply chain structures has become increasingly critical.

This situation has motivated a growing shift toward analysing inter-firm transaction datasets as supply chains using quantitative network approaches.
Currently, detailed datasets on nationwide transaction networks, covering most firms within a nation, make it possible to study supplier-buyer connections on a massive scale. 
For example, the dataset provided by TEIKOKU DATABANK, Ltd. (TDB) contains four million transactions among approximately 650,000 firms, covering the majority of active firms in Japan.
Researchers have studied the structure of inter-firm networks using large-scale datasets \cite{borgatti2009social_network_analysis,bacilieri2025firmlevel_production_networks}.
In most of this work, an inter-firm transaction graph is treated as a single system, using whole-network descriptors, node centralities, or other global properties to characterise its structure and behaviour. 
This `one-graph' view is attractive because it can handle large numbers of firms and provide standardised metrics for comparison across different settings.

\begin{figure}
  \centering
  \includegraphics[width=\linewidth]{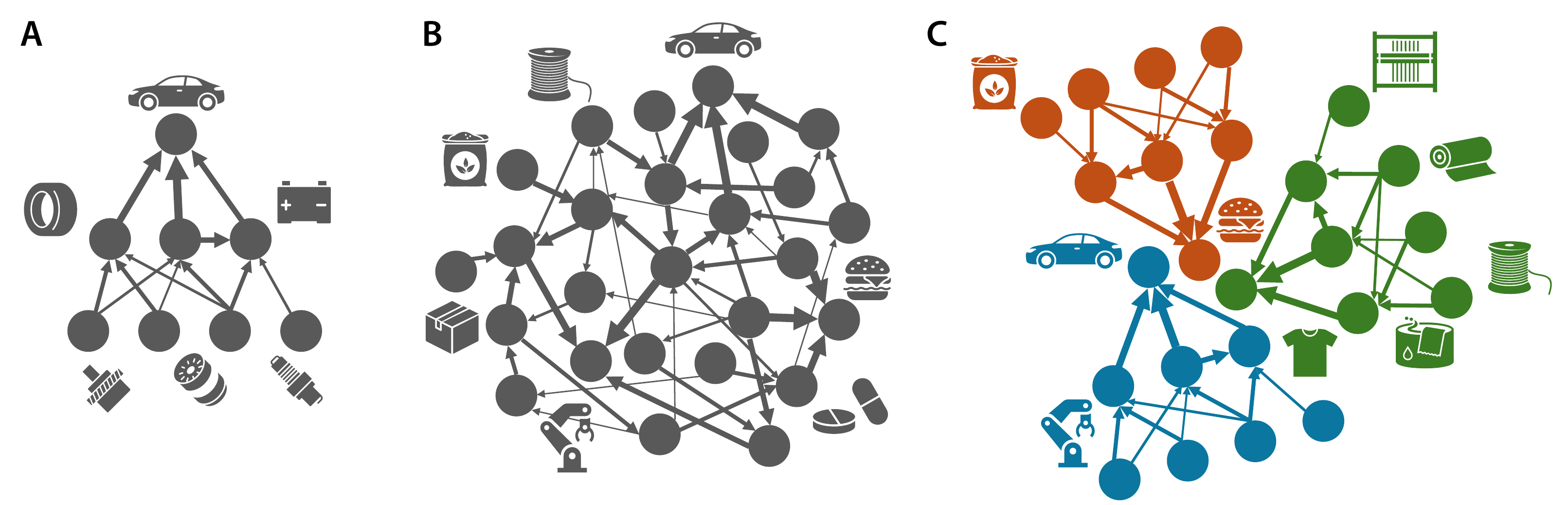}
  \caption{
    \textbf{(A)}
    The concept of supply chain emphasizes a hierarchical, ordered flow  from upstream suppliers through intermediate stages to downstream assemblers.
    \textbf{(B)}
    Real-world inter-firm data rarely resembles an acyclic, flow-hierarchical structure.
    \textbf{(C)}
    We propose flow-Hierarchical clustering to recover the hierarchical tiers and upstream--downstream positioning of individual firms within the clusters.
  }
  \label{fig:illustration}
\end{figure}

However, supply-chain theory implies more than simple `connectivity' \cite{mentzer2001defining_supply_chain,lazzarini2001integrating_supply_chain}.
It emphasises a hierarchical, ordered flow, wherein inputs move from upstream suppliers through intermediate stages to downstream assemblers, distributors, and customers, as shown in Figure \ref{fig:illustration}(A).
A critical limitation is that what we observe in large-scale inter-firm data rarely resembles an acyclic, flow-hierarchical structure.
Instead, as illustrated in Figure \ref{fig:illustration}(B), the full transaction network often appears as an interwoven supply network \cite{ivanov2020viability_intertwined_supply}; for example, a supplier in the food production process can be a buyer of products, such as chemical fertilisers and agricultural machinery.
Such overlapping supply chains and cycles entangle flows, obscuring the pyramidal production-and-assembly pathways that concepts like `tiers' and `ordering' presume.
Without appropriate partitioning of inter-firm transaction networks involving millions of firms, the flow-hierarchical structures that supply-chain concepts are meant to capture cannot be revealed.

To address this mismatch between observable transaction networks and the flow-hierarchical structures implied by supply-chain concepts, we propose the \emph{flow-Hierarchical Community Network Extraction} (f-HiCoNE) framework, which extracts supply-chain-relevant clusters from an entire transaction network using combinatorial Hodge decomposition, which separates the observed flow into acyclic (gradient flow driven by scalar potential) and cyclic components, as explained later.
f-HiCoNE proceeds in three steps: (i) quantifying the connection strength between firms with the acyclic flows, (ii) partitioning the graph based on the strength of acyclic connectivity, and (iii) ordering firms within each extracted cluster using locally recalculated scalar potentials.
This approach recovers the hierarchical tiers and upstream-downstream positioning of individual firms (Figure~\ref{fig:illustration}C).
We apply f-HiCoNE framework to the TDB nationwide transaction network, yielding 27 distinct clusters.
Most clusters exhibit a strong flow-hierarchical organisation, with transaction flows well described by the local scalar potential, effectively revealing the supply chain pathways.
Empirically, these clusters tend to be industrially specialised and geographically concentrated.
By combining cluster membership with local scalar potential, firms can identify their absolute positioning within the broader supplier-buyer hierarchy.

Instead of applying a global hierarchy to the entire transaction network, our method finds upstream-downstream hierarchies at the cluster level. While, in relevant previous works, Kichikawa et al. [25] used circular-flow components and de Jonge et al. [26] extracted a global DAG-like backbone, f-HiCoNE's local scalar potential clarifies the supplier-buyer order within specific supply-chain clusters. Firms can obtain the group including them from a large and complex interwoven network and the position within it. Our approach offers a `map' that shows their locations in the surrounding environment.

The remainder of this paper is organised as follows: Section~\ref{sec:literature} reviews related work and positions our contribution within the existing literature. Section~\ref{sec:methods} details the proposed framework and algorithm, including the definition of combinatorial Hodge decomposition, data description, and preprocessing steps. Section~\ref{sec:results} presents case study results and evaluates the internal flow-hierarchy, industrial relevance, and geographic characteristics of the extracted clusters. Finally, Section~\ref{sec:discussion} summarises the findings and discusses study limitations and challenges.

\section{Related works}
\label{sec:literature}
In this section, we review related studies, beginning with the conceptual evolution of the supply chain and clarify the position of our contribution within the existing literature.

Conceptualisation of the supply chain has expanded to encompass all activities related to the transformation and flow of goods--from raw materials to end users--along with the corresponding information flows \cite{handfield1999introduction_supply_chain}. Mentzer et al. \cite{mentzer2001defining_supply_chain} offer a widely accepted definition, describing a supply chain as `a set of three or more entities (organizations or individuals) directly involved in the upstream and downstream flows of products, services, finances, and/or information from a source to a customer'. This definition extends beyond dyadic relationships to characterise a hierarchical network structure comprising multiple tiers. Within this hierarchy, suppliers are stratified by their proximity to the focal firm. Tier 1 suppliers supply the manufacturer directly, whereas Tier 2 suppliers supply Tier 1, thereby forming a multi-echelon network of dependencies \cite{lambert1998supply_chain_management}. While general systems theory frames supply chains as dynamic systems of interacting organisations \cite{caddy2007supply_chains_their}, recent studies increasingly posit these systems as complex adaptive networks rather than linear chains \cite{choi2001supply_networks_complex}. This perspective shifts the analytical focus from process integration to network properties as fundamental determinants of performance and innovation \cite{burt2000network_structure_social}.

The literature on inter-firm transaction networks has evolved from conceptual frameworks that apply social network analysis to supply chains. Foundational review papers \cite{borgatti2009social_network_analysis, bellamy2013network_analysis_supply} have established the theoretical utility of network metrics in understanding supply chain architecture, whereas more recent surveys \cite{carvalho2019production_networks_primer} highlight the shift toward granular, firm-level data to understand aggregate economic fluctuations and systemic risk. Seminal empirical studies using exhaustive Japanese datasets \cite{fujiwara2010largescale_structure_nationwide} have revealed scale-free degree distributions and disassortative mixing. In Europe, Dhyne et al. \cite{dhyne2021trade_domestic_production} analysed Belgian value-added tax (VAT) records to map domestic production networks, demonstrating how sparse connectivity between importers and domestic firms amplifies the local propagation of foreign demand shocks. Recent studies have expanded this scope; for instance, Pichler et al. \cite{pichler2023building_alliance_map} proposed the formation of an international alliance to integrate fragmented firm-level data into a comprehensive global supply network map, arguing that this is essential for addressing systemic risks and managing the green transition. As an application, Carvalho et al. \cite{carvalho2021supply_chain_disruptions, inoue2019firmlevel_propagation_shocks} and Inoue and Todo \cite{inoue2019firmlevel_propagation_shocks} utilised the Great East Japan Earthquake to causally demonstrate how supply chain linkages propagate shocks across distances, collectively confirming that inter-firm networks exhibit nontrivial topologies that significantly influence economic resilience.
Empirical research on inter-firm transaction networks has increasingly used community detection techniques to uncover mesoscale economic structures that transcend traditional sectoral boundaries. For instance, Beckers et al. \cite{beckers2018logistics_clusters_including} applied the Louvain algorithm \cite{blondel2008fast_unfolding_communities} to a dataset of buyer-supplier relations among Belgian logistics firms, aiming to generate a typology of clusters that integrates spatial co-location with relational density to identify distinct `spill-over' and polycentric clusters beyond simple employment concentration. Wiedmer and Griffis \cite{wiedmer2021structural_characteristics_complex} studied the Mergent Horizon database to analyse 21 extended supply chains. For these different datasets, they assessed several well-known network structures, such as scale-free, small-world, and modular structures, and consistently observed the scale-free and modular structures in these datasets.

Recent studies have further extended this analysis by adopting advanced topological and flow-based algorithms to reveal complex sequencing and global production hierarchy. In a notable study of this structural decomposition, Chakraborty et al. \cite{chakraborty2018hierarchical_communities_walnut} analysed the massive Tokyo Shoko Research (TSR) dataset of one million Japanese firms using the Infomap algorithm \cite{rosvall2008maps_random_walks}. They identified a `walnut' structure composed of upstream (IN), downstream (OUT), and Giant Strongly Connected Components (GSCC), revealing that a giant strongly connected core is encased by upstream and downstream shells, with most irreducible communities existing at the second level of the hierarchy. Building on this TSR dataset, Kichikawa et al. \cite{kichikawa2019community_structure_based} aimed to decompose flows into gradient (hierarchical) and circular components (industrial feedback loops) using combinatorial Hodge decomposition. For the isolated circular components, they applied the Infomap algorithm to detect the communities of firms. De Jonge et al. \cite{dejonge2025deriving_production_chains} applied a Restricted Gradient Extraction (RGE) method to the Dutch production network (DPN2018), aiming to project the enterprise network onto a commodity network of 523 nodes to extract a directed acyclic graph (DAG) that isolates the linear gradient-flow of value addition. To address the issue of overlapping communities, Lu and Dong \cite{lu2023gravitationbased_hierarchical_community} developed a method combining the gravity model and hierarchical clustering. Demonstrated using real-world smartphone battery supply chains among 146 firms, this approach uses simulated gravitational forces between nodes to reconstruct multilayered, overlapping community architectures that reflect the intertwined nature of modern industrial outsourcing.

These related works reveal two important facts regarding inter-firm transaction network structures. First, global transaction networks are not acyclic. A significant portion consists of strongly connected components with loops, indicating that no clear global hierarchy exists. Second, community detection analyses demonstrate that these networks possess a modular structure. Firms are organised into distinct groups that are often specialised by industry or region. These observations emphasise the necessity of applying flow-hierarchical clustering to the entire network and imply that the resulting groups possess functional significance.

However, a significant limitation of previous approaches is their focus on hierarchical modularity rather than flow-hierarchy. Hierarchical modularity describes a nested structure in which smaller communities are encapsulated within larger ones. Although Kichikawa et al. \cite{kichikawa2019community_structure_based} similarly adopted combinatorial Hodge decomposition to rank firms' upstream-downstream positions, they argued for hierarchical modularity by applying the Infomap algorithm \cite{rosvall2008maps_random_walks} to circular flows, which are orthogonal to upstream-downstream flows. 
De Jonge et al. \cite{dejonge2025deriving_production_chains} proposed representing a network of 523 commodities as a DAG based on combinatorial Hodge decomposition. 
It is analogous to extracting a Minimum Spanning Tree; their objective was to identify the backbone structure of the commodity graph rather than partitioning it. By contrast, we specifically address the flow-hierarchy inherent in supplier-buyer relationships rather than topological modularity and evaluate the flow-hierarchy within individual clusters rather than across the entire network.

\section{Methods}
\label{sec:methods}
This section is organised into three main subsections. First, we describe the dataset analysed in this study and explain how economic transaction records between firms are modelled as a directed network of supplier-buyer relationships. Second, we introduce combinatorial Hodge decomposition as the mathematical foundation for separating an acyclic hierarchical component from cyclical components in a graph. Finally, we detail the proposed f-HiCoNE algorithm.

\subsection{Dataset}
\label{sec:dataset}
We used an inter-firm transaction dataset provided by TDB \cite{tdbweb}, a leading credit research company with offices across all prefectures in Japan. TDB collects transaction data during corporate credit research because the financial condition of business partners directly influences a firm's creditworthiness. In this research, firms disclose information about their suppliers and customers. By integrating these Corporate Credit Reports (CCR), a large-scale inter-firm network is constructed; for instance, if Company B lists a supplier or customer also reported by Company A, their respective ego networks are linked. Importantly, this dataset includes inferred transaction amounts \cite{tamura2015,tamura2018,JP_B9_6860731,JP7097500B1}. We used the latest available transaction records over a three-year period (2022/01/01 to 2024/12/31), denoted as a square matrix $M$. The element $M_{ij}$ represents the transaction amount from supplier $j$ to buyer $i$, indicating that goods or services flow from $j$ to $i$, while payments flow from $i$ to $j$. 

Using the transaction matrix $M$, we construct a network of supplier-buyer relationships, in which edges are assigned between firms with transactions $M_{ij} > 0$ or $M_{ji} >0$. To quantify the net directionality of transactions between firms $i$ and $j$, we define flow $Y_{ij}$ based on the asymmetry of transaction volumes as follows:
\begin{align}
  Y_{ij}  =  \frac{M_{ij} - M_{ji}}{M_{ij} + M_{ji}}.
\end{align}
The matrix $Y$ is skew-symmetric ($Y_{ij}=-Y_{ji}$), where a negative sign indicates flow in the opposite direction. 
This formula generalises the relationship structure: for strictly unidirectional transactions (where either $M_{ij}$ or $M_{ji}$ is zero), it yields $|Y_{ij}| = 1$. 
In bidirectional cases, $Y_{ij}$ approaches 1 when the flow $i\rightarrow j$ dominates ($M_{ij}\gg M_{ji}$). 
Note that the ratio is undefined ($0/0$) if no transactions occur ($M_{ij}=M_{ji}=0$); in this case, no edge is assigned. This is distinct from the zero-weight situation ($Y_{ij}=0$), which arises from perfectly balanced bidirectional trade ($M_{ij}=M_{ji}>0$). 
We focus on the largest weakly connected component of the network, which contains approximately 99.63 percent of firms in the dataset. Finally, we obtain a network with edge flows $Y$ derived from $M$, which has 629,098 firms and 4,247,020 links.

\subsection{Combinatorial Hodge Decomposition}
\begin{figure}
  \centering
  \includegraphics[width=\linewidth]{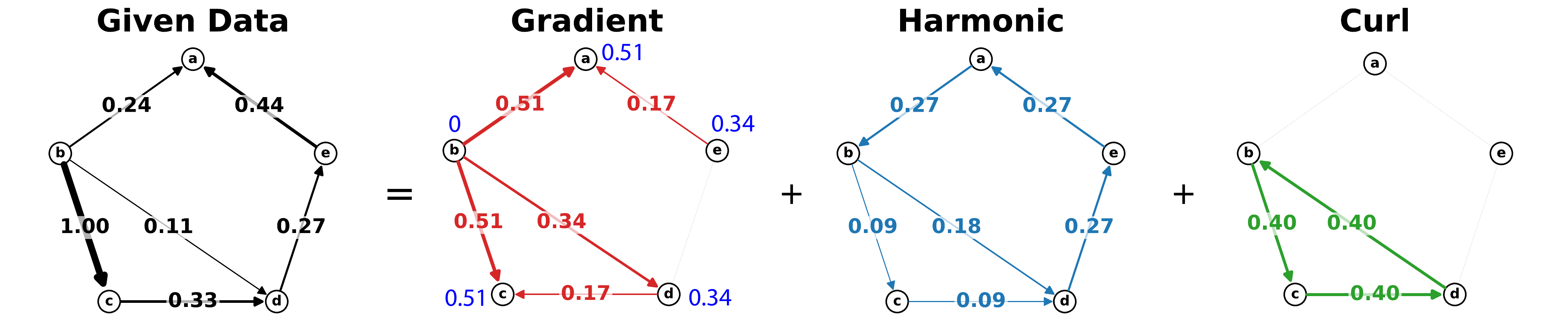}
  \caption{Combinatorial Hodge decomposition isolates the acyclic component from a given inter-firm transaction network,
    which is described by gradient of scalar potential of firms.
    Scalar potential of each firm is denoted by blue texts and the difference between firms gives the gradient flows between them.
  }
  \label{fig:hodge}
\end{figure}

Here, we outline the combinatorial Hodge decomposition formulated by Jiang et al. \cite{jiang2011statistical_ranking_combinatorial}. 
Let $G\left(V,E\right)$ denote an undirected graph with a vertex set $V$ and an edge set $E$, where $N$ is number of vertices.
Let $Y$ be an $N \times N$ matrix representing flows on $G$. 
We assume that $Y$ is skew-symmetric, satisfying $Y_{ij}=-Y_{ji}$. 
In this study, we applied this framework to the supplier-buyer network characterised by edge flow $Y$ defined in the previous section.

The combinatorial gradient, curl, and divergence are defined as
\begin{align*}
  (\text{grad}\, s)(i, j) &=  s_j-s_i \quad \text{for $\{i,j\} \in E$} , \\
  (\text{curl}\, Y)(i, j, k) &= Y_{ij} + Y_{jk} + Y_{ki}\quad \text{for  $\{i,j,k\}$}: \{i, j\}, \{j, k\}, \{k, i\} \in E ,\\
  (\text{div}Y)(i) &= \sum_{j \text{ s.t. } \{i,j\} \in E} Y_{ij},
\end{align*}
where $s$ denotes the scalar potential of the vertices. The space of edge flow $\mathcal{Y}$ is orthogonally decomposed into the images and kernels of these operators as follows:
\begin{align*} 
  \mathcal{Y}  = \text{im}(\text{grad})  \oplus \text{ker}(\Delta_1) \oplus  \text{im}(\text{curl}^*), %
\end{align*}
where $\text{ker}(\Delta_1) = \text{ker}(\text{curl}) \cap \text{ker}(\text{div})$ corresponds to the harmonic component, and $\text{curl}^*$ is the adjoint of the curl operator. 
As illustrated in Figure ~\ref{fig:hodge}, any given edge flow $Y$ on $G\left(V,E\right)$ is uniquely decomposed into three components: gradient, curl, and harmonic flow.

Gradient flow represents the acyclic component of the observed flow, which is determined by the difference in the scalar potential of firms. 
When a firm $j$ has a higher potential than a connected firm $i$ (i.e.\ $s_j>s_i$), the gradient flow yields a positive value of $s_j-s_i$. 
As defined above, firm $j$ acts as the supplier and firm $i$ acts as the buyer. 
Therefore, firms can be sorted in descending order from upstream to downstream based on their scalar potential. 
Moreover, owing to the global consistency of the gradient flow, the accumulated flow between any two firms is path-independent. 
For example, the gradient flow along the path $b\rightarrow c$ (+0.51) equals that along an alternative path $b\rightarrow d\rightarrow c$ (0.34 + 0.17 = 0.51). 
Consequently, scalar potential serves as a consistent measure of a firm's upstream-downstream position.

The remaining components are circular. Owing to the orthogonality of the decomposition, both harmonic and curl flows are divergence-free. Therefore, incoming and outgoing flows are balanced at every node, forming closed loops. Curl flows form local loops among three firms, whereas harmonic flows form larger cycles involving more than three firms. Within these circular flows, no distinct upstream or downstream hierarchy exists.

The scalar potential $s$ is defined as the solution to the following optimisation problem of least squares:
\begin{align}
  \min_s \sum_{\{i,j\} \in E} \left[ \text{(grad $s$)}(i,j) - Y_{ij} \right]^2
  =
  \min_s \sum_{\{i,j\} \in E} \left[ (s_j - s_i) - Y_{ij} \right]^2. \label{eq:optimization}
\end{align}
The above equation represents a problem of finding the closest point to the given data $Y$ in the subspace of the edge flow and can be solved by an $l_2$-projection of $Y$ onto im(grad) \cite{jiang2011statistical_ranking_combinatorial}. 
With a Euclidean inner product in space $\mathcal{Y}$, $\langle X,Y\rangle = \sum_{ \{i,j\} \in E}  X_{ij}Y_{ij}$, the normal equation is given by
\begin{align}
  \Delta_0 s  = - \text{div} Y,  \label{deltas}
\end{align}
where $\mathrm{\Delta}_0$ is the graph Laplacian of the undirected graph $G\left(V,E\right)$. Finally, the potential s is given by the minimal-norm solution of Equation \eqref{deltas}
\begin{align}
  s  = - \Delta_0^{\dagger} \text{div} Y,  \label{eq:potential}
\end{align}
where $\dagger$ denotes the Moore-Penrose inverse.

Decomposition quantifies the extent to which a network with edge flow Y fits the flow-hierarchical structure. 
Based on the least-squares optimisation \eqref{eq:optimization}, the ratio of the gradient component $R^2$ \cite{haruna2016hodge_decomposition_information,aoki2022urban_spatial_structures} is given by
\begin{align}
  R^2 = \frac{\sum_{\{i,j\}\in E} \left[ \text{(grad $s$)}(i,j)\right]^2}{\sum_{\{i,j\}\in E} Y_{ij}^2}.
\end{align}
This metric corresponds to the `coefficient of determination' in statistics and measures the explanatory power of the scalar potential.

\subsection{f-HiCoNE algorithm}

\begin{figure}
  \centering
  \includegraphics[width=\linewidth]{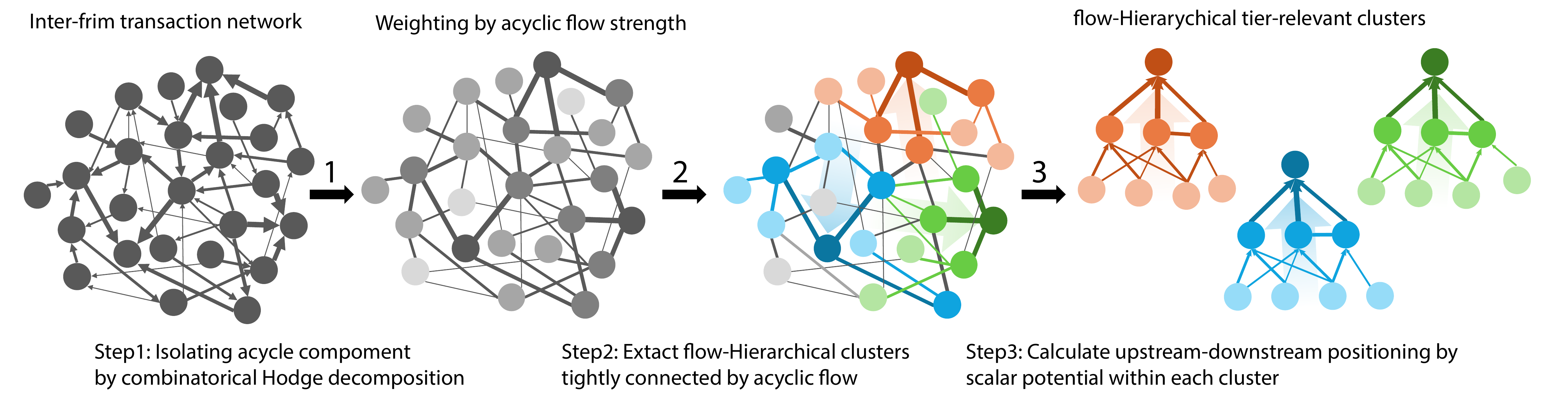}
  \caption{
    \textbf{The flow-Hierarchical Community Network Extraction (f-HiCoNE) algorithm.}
    The process consists of three stages: 
    \textbf{(Step 1)} Quantifying the connection strength of firms with the acyclic flow to define new edge weights $w_{ij}$;
    \textbf{(Step 2)} Partitioning the graph $G(V,E,W)$ into clusters dominated by the acyclic flow using these weights; and
    \textbf{(Step 3)} Calculating the scalar potential within each extracted cluster to determine the firms' hierarchical positions. 
  }
  \label{fig:algorithm}
\end{figure}

f-HiCoNE is a heuristic algorithm that can extract flow-hierarchical clusters from entire network data and is scalable to large datasets comprising millions of firms. 
Figure \ref{fig:algorithm} illustrates this algorithm, which proceeds in three steps.

Step 1 isolates the acyclic component of the supplier-buyer network using combinatorial Hodge decomposition and quantifies the connection strengths via the gradient flow:
\begin{align}
  w_{ij} = \lvert \text{(grad $s$)}(i,j) \rvert.
\end{align}
Applying combinatorial Hodge decomposition, any flow can be uniquely decomposed into parts whose inner product is zero, much like separating a signal (acyclic component) into completely distinct non-interfering channels (cyclic components).
These weights $W$ identify firms strongly connected by the gradient component obtained this decomposition. By replacing the original edge flows $Y$ with $W$, we obtain an undirected, weighted graph $G\prime\left(V,E,W\right)$.

Step 2 extracts flow-hierarchical clusters from $G\prime\left(V,E,W\right)$. Although various clustering methods are applicable to our framework, we employ the Infomap algorithm \cite{rosvall2008maps_random_walks} because of its computational efficiency in analysing nearly one million firms.

Step 3 determines the upstream-downstream positioning of individual firms within the clusters using scalar potentials derived from the original edge flow $Y$.
This potential is recalculated locally within each cluster, which is distinct from the global potential derived in Step 1. 
Subsequently, we used the $R^2$ metric to verify the flow-hierarchy of the extracted clusters.

\section{Results}
\label{sec:results}
\begin{figure}
  \centering
  \includegraphics[width=\linewidth]{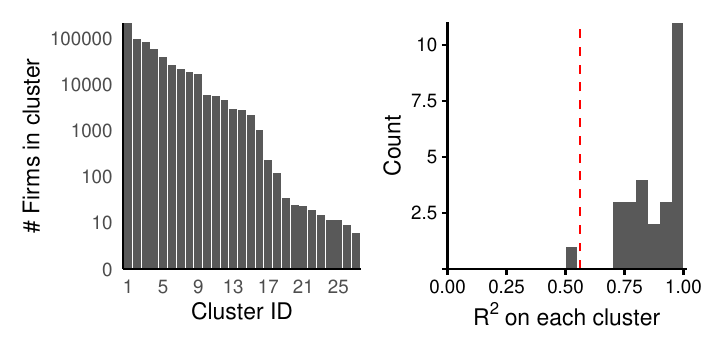}
  \caption{
    \textbf{Characteristics of extracted flow-hierarchical clusters.}
    \textbf{(A)} Cluster sizes (number of firms).
    \textbf{(B)} Distribution of the coefficient of determination, $R^2$. The red vertical line indicates the $R^2$ of the entire network.
  }
  \label{fig:cluster_description}
\end{figure}

Applying the f-HiCoNE method to the inter-firm transaction dataset (Section \ref{sec:dataset}) yielded 27 distinct clusters. As shown in Figure \ref{fig:cluster_description}(A), the largest cluster comprised approximately 220,000 firms, whereas nine other major clusters each contained over 10,000 firms. Eighteen clusters exceeded 100 firms in size.

These clusters generally exhibited a stronger flow-hierarchy than the entire network. Figure \ref{fig:cluster_description}(B) shows the $R^2$ distribution of clusters against the network-wide baseline (red vertical line). With one exception ($R^2\sim0.53$), all clusters achieved a higher $R^2$ than the baseline. The outlier was the second-largest cluster, comprising approximately 98,000 firms. This deviation is expected because large, complex transaction networks often contain tangled webs of firms with numerous feedback loops, which remain after partitioning flow-hierarchical clusters and can suppress the gradient component even after partitioning. Conversely, the remaining clusters demonstrated consistently high $R^2$ values ($\in\left[0.713,1\right]$), confirming that the gradient component is dominant and that the potential-driven hierarchical structure, that is, gradient flows, well describes internal flows. Consequently, for these clusters, the scalar potential serves as a reliable indicator of the upstream-downstream positioning of firms.

\begin{figure}
    \centering
    \includegraphics[width=\linewidth]{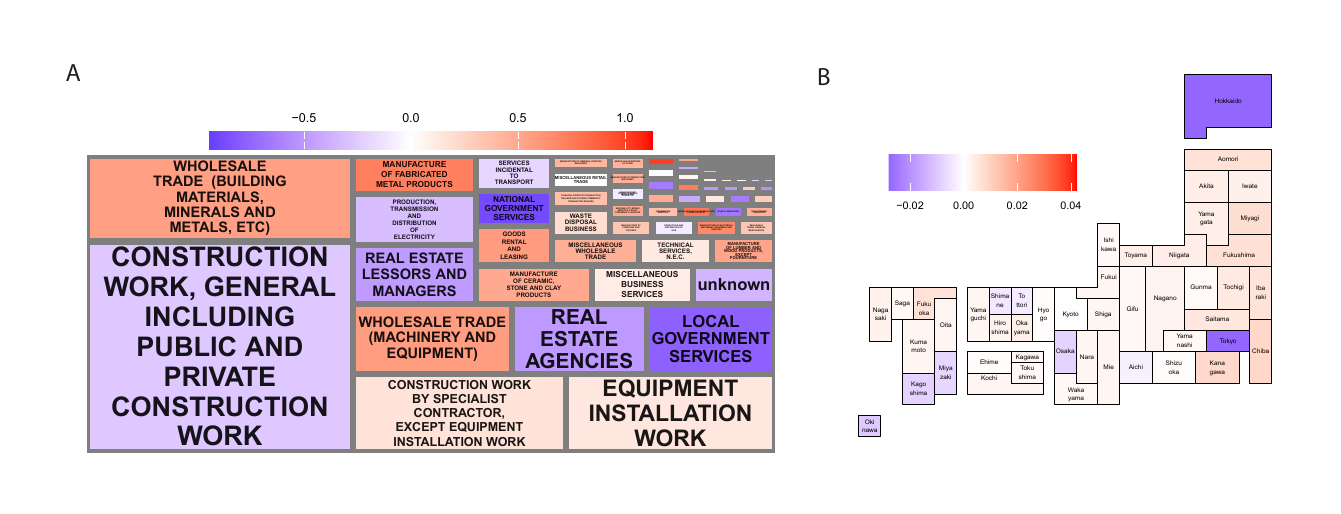}
    \caption{
  \textbf{Industrial and geographical characteristics of Cluster 1.}
  \textbf{(A)} Industrial composition visualised as a treemap, grouped by JSIC 2-digit sectors.
  Rectangle area is proportional to the sector's total transaction amount within the cluster, and colour represents the median of the scalar potentials (red: upstream/high potential; blue: downstream/low potential).
  \textbf{(B)} Relative geographical density of firm headquarters by prefecture in the cluster, compared to the baseline of all firms in the dataset.
}
    \label{fig:cluster01}
\end{figure}

We then examined the industrial composition and geographical distribution of the extracted clusters. Figure \ref{fig:cluster01}(A) summarises the industrial composition of the largest extracted cluster (Cluster 1) using the Japan Standard Industrial Classification (JSIC) at the 2-digit level. The figure is presented as a treemap, where each rectangle corresponds to an industrial sector. The area of the rectangle is proportional to the total transaction amount, $\sum_{j \in C} M_{ij}+M_{ji}$, where $C$ denotes the set of firms in that sector. The sector-level median scalar potential is encoded in colour, with a red-to-blue gradient representing higher (upstream) to lower (downstream) potentials. As shown in Figure \ref{fig:cluster01}(A), ‘Construction work, general including public and private construction work’ occupied the largest areas and exhibited negative scalar potential, placing them on the downstream (consumer) side of Cluster 1. Real-estate related sectors, such as ‘Real estate agencies’ and ‘real estate lessors and managers’ also showed similar negative potentials, suggesting downstream positioning that could reflect direct provision to entities such as construction work. Local and national government services exhibited more pronounced negative potentials, suggesting that the construction work and real estate could be demanded by government. By contrast, sectors with positive potentials included upstream-oriented wholesale and manufacturing activities, such as ‘Wholesale trade (building materials, minerals and metals, etc.)’, ‘Wholesale trade (machinery and equipment)’, ‘Manufacture of fabricated metal products’, and ‘Manufacture of ceramic, stone and clay products’. ‘Construction work by specialist contractor, except equipment installation work’ and ‘Equipment installation work’ sectors appeared at the intermediate positions (near-zero potentials). This configuration suggests a distinct supply chain for public infrastructure, such as roads, bridges, tunnels, and buildings by governments, where construction firms procure materials from upstream wholesalers and manufacturers to meet government demand. This cluster also includes parallel pathways. Sectors such as ‘Services incidental to transport’ and ‘Production, transmission and distribution of electricity’ appeared on the consumer side, whereas ‘Waste disposal business’ was positioned upstream, which could point to parallel pathways integrated into the same ecosystem. Figure \ref{fig:cluster01}(B) depicts the geographical distribution of firms in Cluster 1 at the prefectural level. Using headquarters locations, prefectural densities were computed for Cluster 1. We also calculated the baseline density from all firms in the analysed transaction records. The figure shows the resulting density by prefecture relative to the baseline density. Figure \ref{fig:cluster01}(B) shows that Cluster 1 had lower relative density in Tokyo and Hokkaido than the baseline, while being relatively concentrated in several prefectures across the Kanto and Tohoku regions (including Chiba, Kanagawa, Saitama, Ibaraki, Fukushima, Miyagi, and Aomori). This geographic pattern could be interpreted as indicating regionally distributed public sector-related networks rather than activity concentrated in a single metropolitan core.

\begin{figure}
    \centering
    \includegraphics[width=\linewidth]{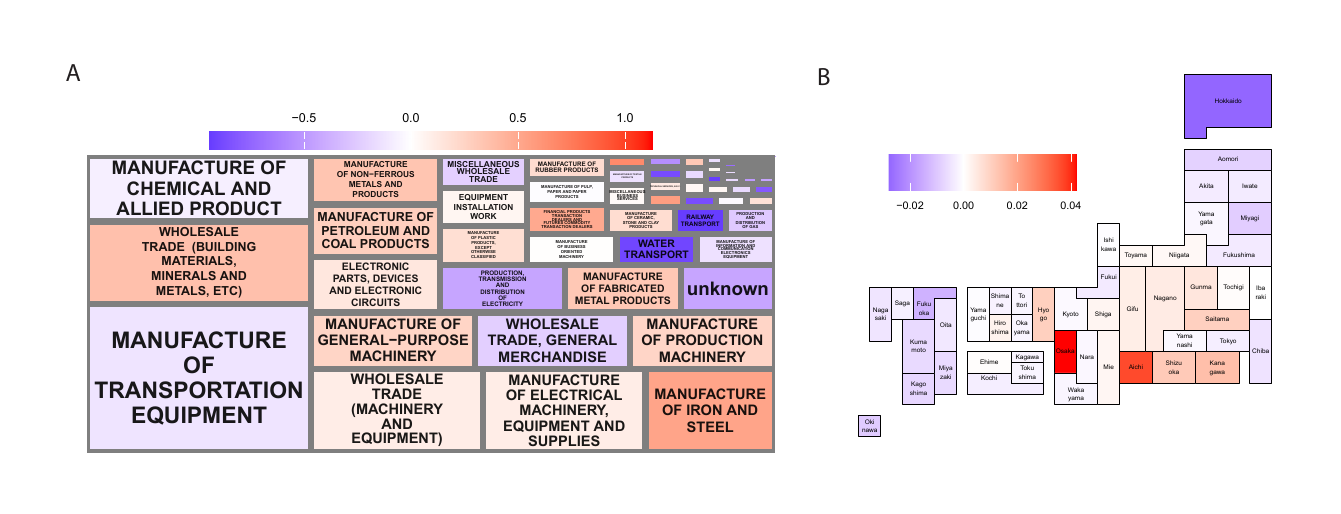}
    \caption{
    \textbf{Industrial and geographic characteristics of Cluster 2.}
    Panels \textbf{(A)}--\textbf{(B)} and the visual encodings are defined as in Fig.~\ref{fig:cluster01}.
    }
    \label{fig:cluster02}
\end{figure}
\begin{figure}
    \centering
    \includegraphics[width=\linewidth]{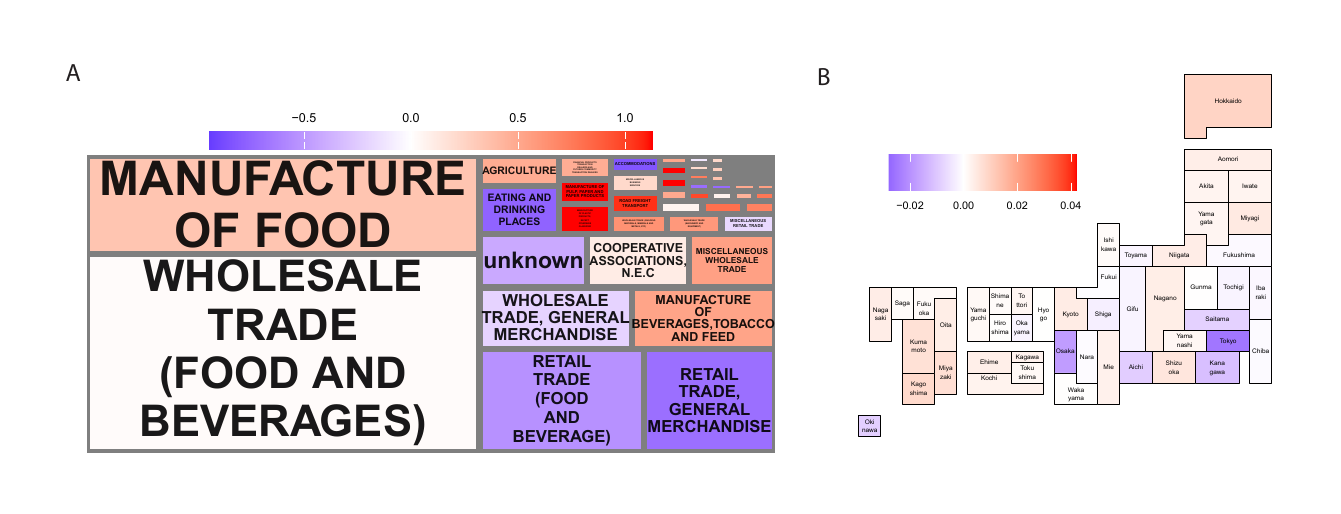}
    \caption{
    \textbf{Industrial and geographic characteristics of Cluster 3.}
    Panels \textbf{(A)}--\textbf{(B)} and the visual encodings are defined as in Fig.~\ref{fig:cluster01}.
    }
    \label{fig:cluster03}
\end{figure}

Figure \ref{fig:cluster02} presents the industrial composition and geographical distribution of the second-largest cluster (Cluster 2), which exhibited a notably lower $R^2$ ($\sim0.53$) than other clusters. As shown in Figure 6(A), the sector ‘Manufacture of transportation equipment’ accounted for the largest share of total transaction amount, and we confirmed the presence of major Japanese automobile manufacturers within this group. As shown in Figure \ref{fig:cluster02}(B), this cluster was geographically concentrated in prefectures that host these key firms: Aichi (Toyota), Osaka (Daihatsu), Shizuoka (Suzuki, Yamaha), Kanagawa (Nissan), and Saitama (Honda). While transportation equipment manufacturers occupied a downstream position (negative potential), sectors such as ‘Railway transport’ and ‘Water transport’ were positioned even further downstream with highly negative potentials, possibly serving as the industrial clients of transportation equipment. Other sectors, such as ‘Manufacture of chemical and allied products’ and ‘Production, transmission, and distribution of electricity’, were located downstream with negative potentials. Conversely, upstream suppliers include ‘Manufacture of iron and steel’, ‘Manufacture of production machinery’, ‘Manufacture of fabricated metal products’ ‘Manufacture of non-ferrous metals and products’, ‘Manufacture of petroleum and coal products’, and ‘Wholesale trade (building materials, minerals, metals, etc.)’ and other manufacturers. Intermediate sectors (positive potential near zero) comprised ‘Wholesale trade (machinery and equipment)’, ‘Manufacture of electrical machinery, equipment and supplies’, and ‘Electronic parts, devices and electronic circuits’. These findings suggest that Cluster 2 encompasses multiple interwoven production pathways: one flowing from raw materials (iron and steel) through intermediate machinery to electronic components and transportation equipment and another from petroleum/coal to chemical products or electricity generation. The partial overlap of these distinct supply chains likely contributes to the observed lower $R^2$ value, reflecting a more complex and less strictly hierarchical network structure.

Next, we examined the industrial composition and geographical distribution of Cluster 3. As shown in Figure \ref{fig:cluster03}(A), this cluster was primarily characterised by food-related production and distribution. The largest sectors, ‘Wholesale trade (food and beverages)’ and ‘Manufacture of food’, exhibited potentials close to zero or slightly upstream. Downstream activities with negative potentials included ‘Retail trade (food and beverage)’, ‘Retail trade, general merchandise’, and ‘Eating and drinking places’, which serve as the consumer-facing endpoints of the cluster. Conversely, upstream sectors with positive potentials comprised primary and supporting activities, such as ‘Agriculture’ and ‘Manufacture of beverages, tobacco and feed’, alongside logistics-related sectors (e.g. ‘Road freight transport’ and ‘Manufacture of pulp, paper and paper products’) that could facilitate physical distribution within the cluster. Taken together, this potential ordering is consistent with a food-and-beverage supply chain wherein agricultural and processed-food outputs flow through wholesale and logistics channels to the retail and food-service sectors. Within this structure, ‘Cooperative associations, N.E.C.’ and ‘wholesale trade, general merchandise’ plausibly operated near the middle of the network as coordinating entities linking producers to downstream channels. Regarding the geographical distribution, Figure \ref{fig:cluster03}(B) reveals a relatively diffuse pattern for Cluster 3, exhibiting only modest deviations in prefecture-level density. Notably, the concentration of firms was lower in Tokyo and Osaka, whereas most other regions remained close to the national baseline, suggesting limited geographic specialisation.

Further analyses of the industrial composition and geographical distribution of the remaining clusters are provided in the Appendix.

\section{Discussion}
\label{sec:discussion}
In this study, we developed a computational framework to detect latent supply chain structures embedded in large-scale transaction networks. By applying combinatorial Hodge decomposition, our method partitions the entire network into clusters, where the flow-hierarchical structure from suppliers to buyers can be captured by the scalar potential. The coefficient of determination confirms that these extracted clusters were dominated by acyclic gradient components. We also empirically validated the clear hierarchical industrial relationships between upstream suppliers and downstream buyers, alongside distinct geographical localisation. These results demonstrate that maximising the gradient flow ratio within clusters is an effective strategy for disentangling complex economic networks into interpretable functional units.

Our approach diverges from prior network analyses by explicitly targeting the flow-hierarchy inherent in the concept of supply chain rather than topological modularity. While several previous studies have utilised combinatorial Hodge decomposition, they often focus on nested community structures as hierarchical modularity by analysing circular flows orthogonal to the upstream-downstream gradient \cite{kichikawa2019community_structure_based}. Similarly, efforts to extract backbone structures, such as transforming commodity networks into DAGs \cite{dejonge2025deriving_production_chains}, aim to simplify global connectivity rather than partition networks into functional units. By contrast, by evaluating the flow-hierarchy within individual clusters, our method decomposes a complex network into distinct supply chains.

Revealing structural embeddedness in supply--chain management \cite{choi2008structural_embeddedness_supplier} is a possible application of the proposed method.
With uncertainty growing due to conflicts, disasters, and pandemics, firms need to assess supply chain risks.
However, it is difficult to find the actual range of a supply chain and a firm's exact position within it.
The proposed method offers a practical way for corporate managers to map their business environment and locate their position within a massive inter-firm network.
Instead of looking at a confusing web of countless companies,
they can now identify their specific group, see which firms are closely related with similar positions, and find the true upstream sources.
This information on structural embeddedness could help them translate these analytical insights into managerial practice.

This study opens a new research avenue focused on flow-hierarchy clustering in supply chain analysis. Substantial room for methodological development remains. Future work should incorporate overlapping community detection because fundamental industries often serve multiple supply chains simultaneously. For instance, steel manufacturers produce base metals essential to a wide range of sectors; conceptually, these firms should be evaluated as belonging to multiple industrial groups rather than a single cluster. Additionally, while our heuristic algorithm successfully extracts communities with high gradient flow ratios, it lacks a theoretical guarantee of maximisation. Developing a scalable optimisation framework that mathematically ensures the maximisation of gradient components represents a major challenge shared with general community detection, highlighting a key direction for future research.

\bigskip
\noindent\textbf{Acknowledgements}
We are deeply grateful to Takaya Ohsato and Shota Fujishima for their helpful discussions.
\printbibliography

\clearpage
\appendix

\section{Industrial and geographic characteristics of extracted clusters}
In the main text, we have summarized the industrial and geographical characteristics for the largest three clusters.
In this section, we summarize the remaining clusters.

\begin{figure}[htp]
    \centering
    \includegraphics[width=\linewidth]{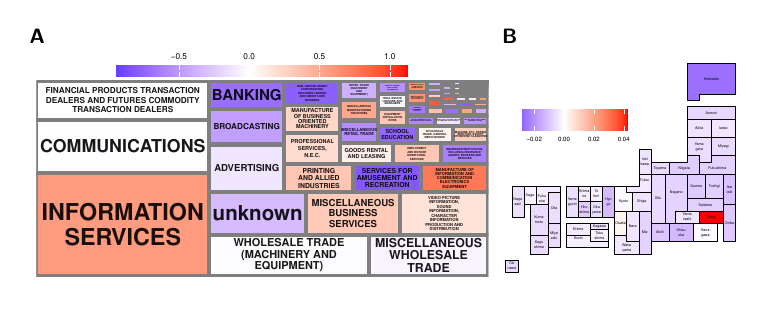}
    \caption{
    Industrial and geographic characteristics of Cluster 4.
    Panels \textbf{(A)}--\textbf{(B)} and the visual encodings are defined as in Fig.~1.
    }
    \label{fig:cluster04}
\end{figure}

In Figure~\ref{fig:cluster04}(a), "Information services" is the largest in scale and dominates the upstream end.
"Manufacture of information and communication electronics equipment" is located even further upstream.
Professional and business-support sectors lie in the mid-to-upstream range, including "Printing and allied industries" and "Video picture information, sound information, character information production and distribution".
Intermediate positions (near-zero potentials) include "Communications", "Financial products transaction dealers and futures commodity transaction dealers", and "Wholesale trade (machinery and equipment)".
Mass-media distribution activities such as "Broadcasting" and "Advertising" appear in the mid-to-downstream range, while finance and consumer-facing media and entertainment sectors such as "Banking" and "Services for amusement and recreation" lie further downstream.
Consequently, this configuration suggests a value chain in which upstream ICT manufacturing and information-service production are linked via content-related production and distribution to downstream media consumption and entertainment.
The downstream placement of financial services, particularly banking, may reflect their association with end-user demand, consumer-facing transactions, or settlement-related functions.
Figure~\ref{fig:cluster04}(b) shows a pronounced concentration in Tokyo, whereas most other prefectures remain close to the baseline level or exhibit lower relative density.
This geographic concentration is consistent with the cluster's orientation toward mass-audience advertising.

\begin{figure}[htp]
    \centering
    \includegraphics[width=\linewidth]{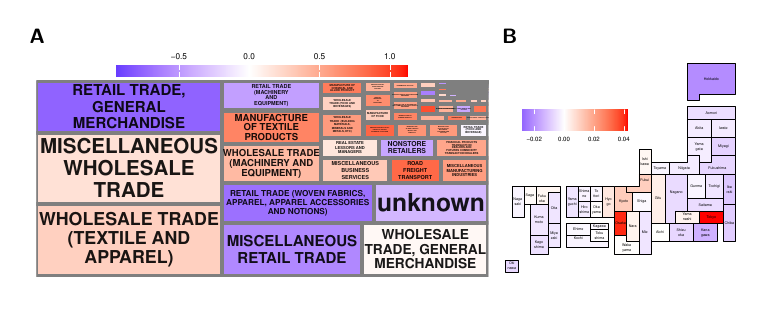}
    \caption{
    Industrial and geographic characteristics of Cluster 5.
    Panels \textbf{(A)}--\textbf{(B)} and the visual encodings are defined as in Fig.~1.
    }
    \label{fig:cluster05}
\end{figure}

We can find textile and apparel wholesale and production activities on the upstream side in Figure~\ref{fig:cluster05}(a).
For example, 'Manufacture of textile products' is accompanied by wholesale sectors such as 'Wholesale trade (textile and apparel)' and 'Wholesale trade (machinery and equipment)'.
Logistics also appear further upstream, most notably 'Road freight transport'.
The flow passes through intermediate positions such as 'Wholesale trade, general merchandise' that serve a distribution function.
From there, it reaches consumer-facing distribution channels on the downstream side: 'Retail trade, general merchandise,', 'Retail trade (woven fabrics, apparel, apparel accessories and notions),' and 'Nonstore retailers'.
This structure points to a process in which raw materials are transported, processed by the textile manufacturing industry, and then delivered to end-customers via wholesale and retail channels.
It is notable that a retail model without physical stores also appears in this cluster.
Figure~\ref{fig:cluster05}(b) indicates that firms in Cluster~5 are relatively concentrated in major metropolitan prefectures, particularly Tokyo and Osaka.
It is natural that this supply chain is concentrated in metropolitan areas, given their large populations, diverse consumer needs, and high demand.

\begin{figure}[htp]
    \centering
    \includegraphics[width=\linewidth]{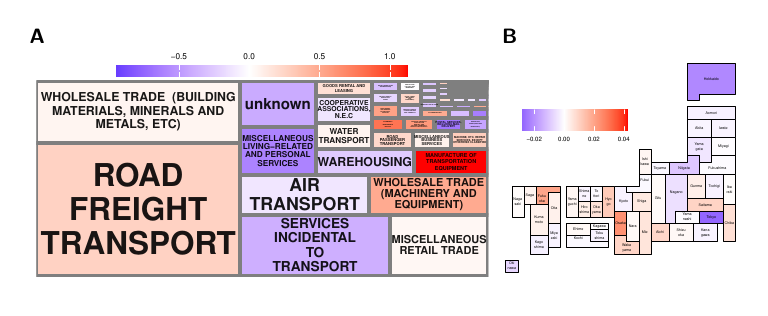}
    \caption{
    Industrial and geographic characteristics of Cluster 6.
    Panels \textbf{(A)}--\textbf{(B)} and the visual encodings are defined as in Fig.~1.
    }
    \label{fig:cluster06}
\end{figure}

In Figure~\ref{fig:cluster06}(a), the most central sector, 'Road freight transport,' is positioned upstream.
'Manufacture of transportation equipment' is smaller in scale but located even further upstream.
'Wholesale trade (building materials, minerals and metals, etc.)' and retail-related activities are located in intermediate positions.
Downstream positions include 'Air transport,' 'Warehousing,' 'Services incidental to transport', and 'Miscellaneous living-related and personal services', among others.
This cluster highlights the role of logistics functions that connect production to downstream services.
In Figure~\ref{fig:cluster06}(b), the relative density is lowest in Tokyo, while it is concentrated around other metropolitan areas.
Logistics functions are needed across a wide area centered on cities, and Tokyo may lack sufficient land for large-scale transportation hubs.

\begin{figure}[htp]
    \centering
    \includegraphics[width=\linewidth]{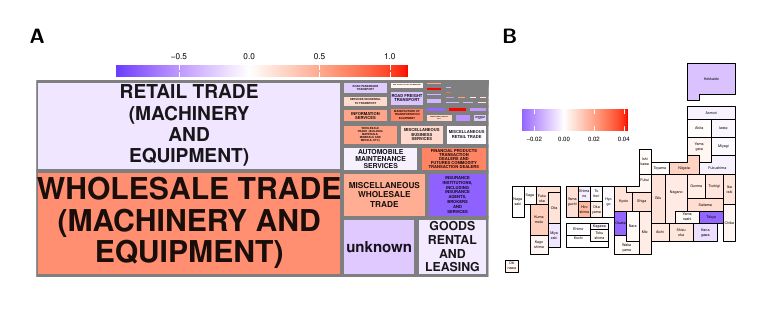}
    \caption{
    Industrial and geographic characteristics of Cluster 7.
    Panels \textbf{(A)}--\textbf{(B)} and the visual encodings are defined as in Fig.~1.
    }
    \label{fig:cluster07}
\end{figure}

Two industries dominate in Figure~\ref{fig:cluster07}(a): wholesale trade and retail trade related to machinery and equipment, and their ordering is consistent with expectations.
'Financial products transaction dealers and futures commodity transaction dealers' appears further upstream, serving as a link between capital and goods.
Financial sectors also appear further downstream, providing support for end users: 'Goods rental and leasing,' 'Insurance institutions, including insurance agents, brokers and services,' and 'Automobile maintenance services' also supports end users.
This structure reflects a lifecycle chain for capital goods.
Figure~\ref{fig:cluster07}(b) indicates that the relative density is low in Tokyo and also lower in Osaka, while the cluster is more prevalent across a range of non-metropolitan prefectures.
This distribution may reflect the need for capital-goods services, including operation and maintenance, to be located close to regional industrial and infrastructure sites.

\begin{figure}[htp]
    \centering
    \includegraphics[width=\linewidth]{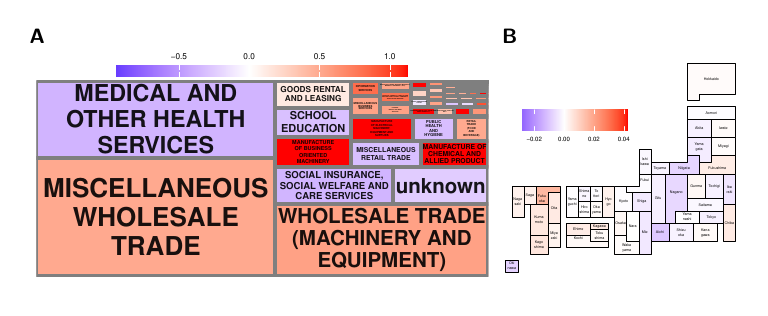}
    \caption{
    Industrial and geographic characteristics of Cluster 8.
    Panels \textbf{(A)}--\textbf{(B)} and the visual encodings are defined as in Fig.~1.
    }
    \label{fig:cluster08}
\end{figure}
In Figure~\ref{fig:cluster08}(a), 'Manufacture of business oriented machinery', 'Manufacture of chemical and allied products', and 'Manufacture of electrical machinery, equipment and supplies' are positioned at the upstream end.
Next, distribution-related sectors such as 'Wholesale trade (machinery and equipment)' and 'Miscellaneous wholesale trade' follow, and these sectors are large in scale.
'Goods rental and leasing' is placed near the intermediate range.
Downstream positions are occupied by institutional and end-use-oriented services such as 'Medical and other health services', 'Social insurance, social welfare and care services', and 'School education'.
This arrangement suggests that goods produced upstream are used by public and welfare-related sectors.
Figure~\ref{fig:cluster08}(b) shows broadly distributed firms across prefectures.
This may reflect the fact that public and social-service institutions are distributed nationwide.

\begin{figure}[htp]
    \centering
    \includegraphics[width=\linewidth]{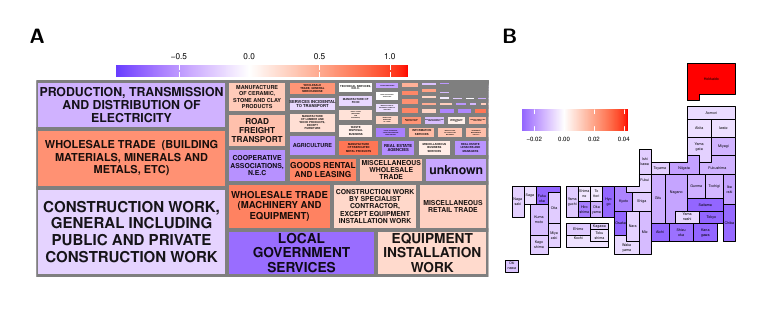}
    \caption{
    Industrial and geographic characteristics of Cluster 9.
    Panels \textbf{(A)}--\textbf{(B)} and the visual encodings are defined as in Fig.~1.
    }
    \label{fig:cluster09}
\end{figure}

'Wholesale trade (building materials, minerals and metals, etc)', 'Wholesale trade (machinery and equipment)' and 'Goods rental and leasing' are positioned clearly upstream in Figure~\ref{fig:cluster09}(a).
'Road freight transport', 'Manufacture of ceramic, stone and clay products', 'Equipment installation work' and 'Construction work by specialist contractor, except equipment installation work' support the preparation of construction-related inputs.
No sector occupies a clear intermediate position in this cluster.
Institutional sectors such as 'Local government services' and 'Cooperative associations, n.e.c' are positioned as end users in this cluster.
Interestingly, Figure~\ref{fig:cluster09}(b) shows a pronounced geographic concentration in Hokkaido, and 'Agriculture' is also present in Figure~\ref{fig:cluster09}(a).
Agriculture is a major industry in Hokkaido.
This cluster may therefore represent a region-specific supply network.

\begin{figure}[htp]
    \centering
    \includegraphics[width=\linewidth]{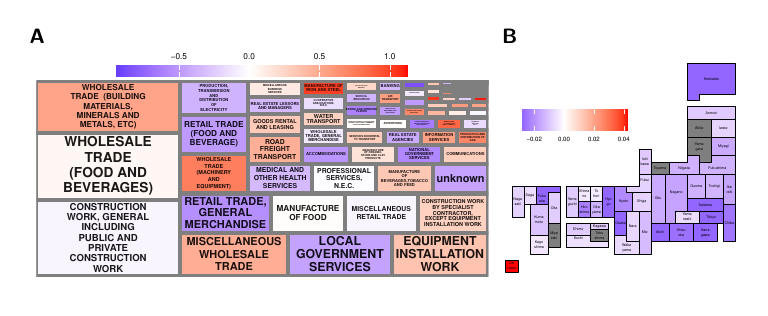}
    \caption{
    Industrial and geographic characteristics of Cluster 10.
    Panels \textbf{(A)}--\textbf{(B)} and the visual encodings are defined as in Fig.~1.
    }
    \label{fig:cluster10}
\end{figure}

Wholesale sectors such as 'Wholesale trade (building materials, minerals and metals, etc)', 'Wholesale trade (machinery and equipment)', and 'Miscellaneous wholesale trade' are positioned on the upstream side in Figure~\ref{fig:cluster10}(a).
'Equipment installation work' and 'Construction work by specialist contractor, except equipment installation work', 'Goods rental and leasing', and 'Road freight transport' also appear on the upstream side.
Food-related supply activities such as 'Wholesale trade (food and beverages)' and 'Manufacture of food' occupy a transitional position.
Consumer-facing channels oriented toward final demand include 'Retail trade, general merchandise', 'Retail trade (food and beverage)', and 'Local government services'.
This positioning implies that the process of converting raw materials into food products and delivering them to consumers intersects with the process of constructing processing facilities.
Figure~\ref{fig:cluster10}(b) indicates a strong concentration in Okinawa, while establishment densities in other prefectures are much lower.
Okinawa is known for its distinctive geographic conditions as an island region, and the clustering method may not fully disentangle industries that are densely co-located and strongly tied to the local context, such as construction and administration and food-related sectors.
Nevertheless, the two processes are related, and the fit remains reasonably good ($R^2=0.772$).

\begin{figure}[htp]
    \centering
    \includegraphics[width=\linewidth]{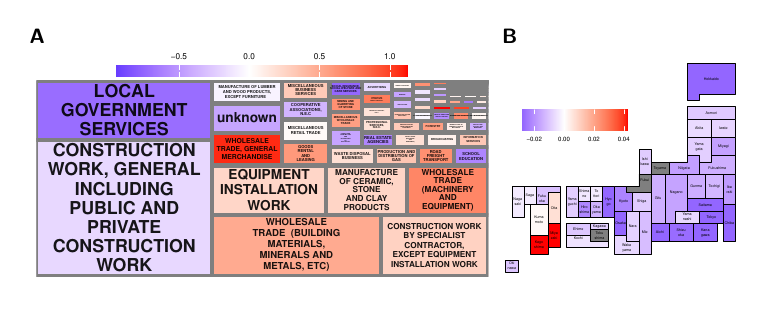}
    \caption{
    Industrial and geographic characteristics of Cluster 11.
    Panels \textbf{(A)}--\textbf{(B)} and the visual encodings are defined as in Fig.~1.
    }
    \label{fig:cluster11}
\end{figure}

In Figure~\ref{fig:cluster11}(a), the words 'Local government', 'Construction', and 'Material' immediately catch the eye, making the cluster easy to interpret at a glance.
On the upstream side, procurement-oriented wholesale sectors such as 'Wholesale trade (machinery and equipment)', 'Wholesale trade (building materials, minerals and metals, etc)', and 'Wholesale trade, general merchandise' are positioned, together with producers of construction-related inputs such as 'Manufacture of ceramic, stone and clay products'.
'Equipment installation work' and 'Construction work by specialist contractor, except equipment installation work' also appear, suggesting a role connecting upstream procurement with downstream on-site execution.
The flow converges on 'Construction work, general including public and private construction work' and 'Local government services'.
Figure~\ref{fig:cluster11}(b) indicates a strong geographic concentration in southern Kyushu, particularly in Kagoshima and Miyazaki.
The relative density is below the baseline in most prefectures, and no establishments are observed in Toyama, Fukui, and Tokushima.
Given this tendency, the cluster appears to represent a region-specific construction and procurement network.

\begin{figure}[htp]
    \centering
    \includegraphics[width=\linewidth]{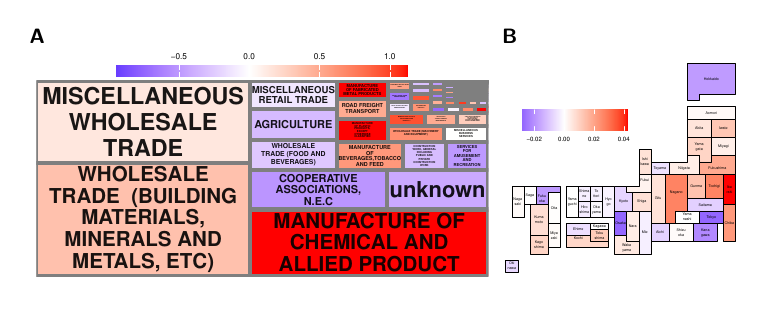}
    \caption{
    Industrial and geographic characteristics of Cluster 12.
    Panels \textbf{(A)}--\textbf{(B)} and the visual encodings are defined as in Fig.~1.
    }
    \label{fig:cluster12}
\end{figure}

In Figure~\ref{fig:cluster12}(a), 'Manufacture of chemical and allied products' is positioned at the upstream end, accompanied by distribution-oriented wholesale sectors such as 'Wholesale trade (building materials, minerals and metals, etc)' and 'Miscellaneous wholesale trade'.
'Agriculture' and 'Cooperative associations, n.e.c', 'Wholesale trade (food and beverages)' and 'Miscellaneous retail trade' appear closer to the consumer side.
This arrangement indicates a structure in which an upstream layer of chemical-product manufacturing is linked to downstream agriculture and distribution, partly through cooperative channels.
Figure~\ref{fig:cluster12}(b) shows lower relative density in major urban areas and comparatively higher presence across multiple regions, including parts of Tohoku, northern Kanto, southern Shikoku, and southern Kyushu.
This distribution suggests geographic proximity between agricultural production sites and upstream input suppliers, in the sense that chemical products used as agricultural inputs and their distribution networks may be concentrated near farming regions rather than in metropolitan areas where corporate headquarters tend to be located.

\begin{figure}[htp]
    \centering
    \includegraphics[width=\linewidth]{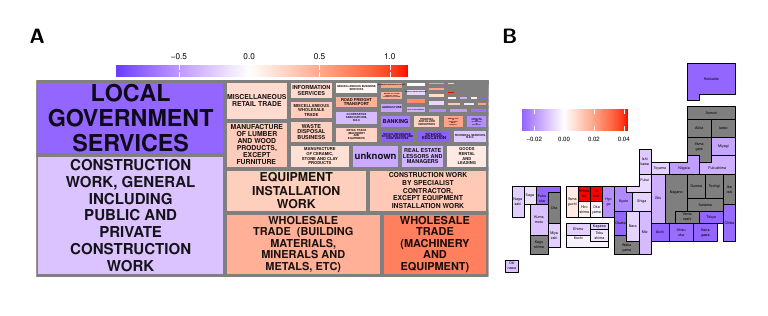}
    \caption{
    Industrial and geographic characteristics of Cluster 13.
    Panels \textbf{(A)}--\textbf{(B)} and the visual encodings are defined as in Fig.~1.
    }
    \label{fig:cluster13}
\end{figure}

In Figure~\ref{fig:cluster13}(a), the words 'Local government', 'Construction', and 'Material' stand out, as in Figure~\ref{fig:cluster11}(a).
Sectors related to the distribution of raw materials and construction equipment are positioned upstream, and the flow converges on public institutions.
Figure~\ref{fig:cluster13}(b) indicates a strong geographic concentration in Tottori and Shimane.
This cluster also captures a region-specific construction and procurement network.
A distinguishing feature is the presence of 'Manufacture of lumber and wood products, except furniture', which suggests that wood products serve as inputs in this cluster.
This region has a relatively high rate of forest coverage.

\begin{figure}[htp]
    \centering
    \includegraphics[width=\linewidth]{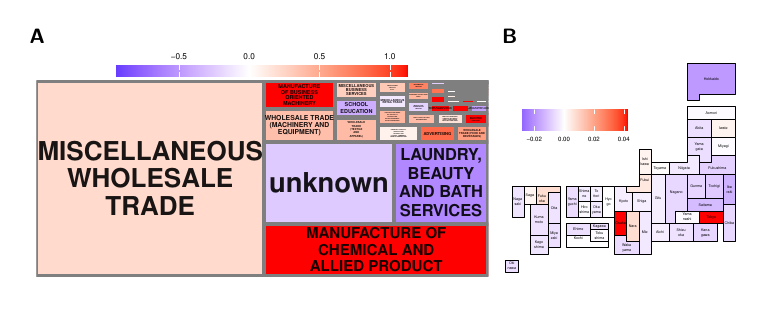}
    \caption{
    Industrial and geographic characteristics of Cluster 14.
    Panels \textbf{(A)}--\textbf{(B)} and the visual encodings are defined as in Fig.~1.
    }
    \label{fig:cluster14}
\end{figure}

'Manufacture of business oriented machinery' and 'Manufacture of chemical and allied products' are positioned at the upstream end, accompanied by distribution-oriented sectors such as 'Wholesale trade (machinery and equipment)' and 'Miscellaneous wholesale trade' in Figure~\ref{fig:cluster14}(a).
Downstream positions include consumer-facing services such as 'Laundry, beauty and bath services', suggesting that part of the downstream demand in this cluster is related to personal services that may rely on chemical products and equipment inputs.
This structure suggests a linkage in which upstream production of chemicals and business machinery connects, through wholesale distribution, to downstream service activities that use these inputs in day-to-day operations.
Figure~\ref{fig:cluster14}(b) indicates a concentration in major metropolitan prefectures, particularly Tokyo and Osaka, while most other prefectures remain close to the baseline level or exhibit lower relative density.
This suggests that manufacturing-related supply and wholesale functions are connected to dense urban service demand.

\begin{figure}[htp]
    \centering
    \includegraphics[width=\linewidth]{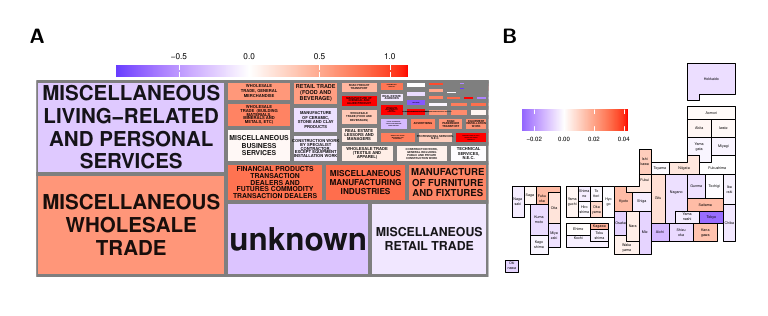}
    \caption{
    Industrial and geographic characteristics of Cluster 15.
    Panels \textbf{(A)}--\textbf{(B)} and the visual encodings are defined as in Fig.~1.
    }
    \label{fig:cluster15}
\end{figure}

In Figure~\ref{fig:cluster15}(a), manufacturing activities such as 'Miscellaneous manufacturing industries' and 'Manufacture of furniture and fixtures' are positioned upstream, accompanied by procurement- and distribution-oriented wholesale sectors such as 'Miscellaneous wholesale trade', 'Wholesale trade (building materials, minerals and metals, etc)', and 'Wholesale trade, general merchandise'.
'Miscellaneous business services' appears around intermediate positions, and consumer-facing activities such as 'Miscellaneous retail trade' and 'Miscellaneous living-related and personal services' appear on the downstream side.
This arrangement shows a structure in which diversified manufacturing is linked to retail and living-related personal services.
Figure~\ref{fig:cluster15}(b) shows that such services are concentrated in urban areas with large populations, such as Tokyo and Osaka.

\begin{figure}[htp]
    \centering
    \includegraphics[width=\linewidth]{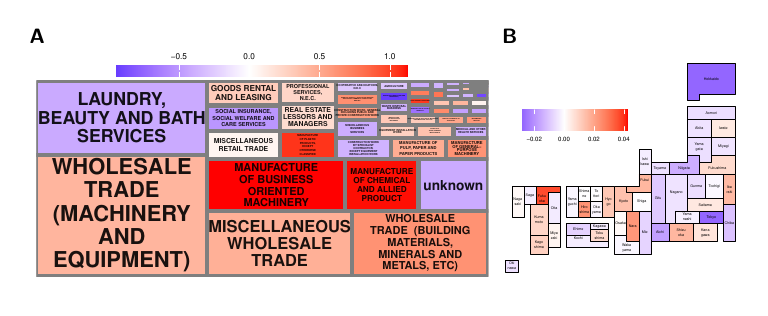}
    \caption{
    Industrial and geographic characteristics of Cluster 16.
    Panels \textbf{(A)}--\textbf{(B)} and the visual encodings are defined as in Fig.~1.
    }
    \label{fig:cluster16}
\end{figure}

The structure of Figure~\ref{fig:cluster16}(a) shows a similar tendency to Figure~\ref{fig:cluster14}(a).
This cluster is also oriented toward living-related personal services, but its geographical pattern differs from that of Cluster~14.
Osaka exhibits average values, and Tokyo shows relatively low levels.
The distribution of high- and low-value prefectures is fragmented nationwide, with a notable concentration of firms in Fukuoka.

\begin{figure}[htp]
    \centering
    \includegraphics[width=\linewidth]{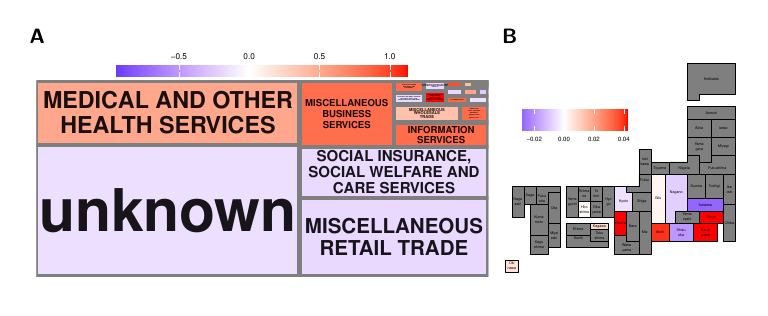}
    \caption{
    Industrial and geographic characteristics of Cluster 17.
    Panels \textbf{(A)}--\textbf{(B)} and the visual encodings are defined as in Fig.~1.
    }
    \label{fig:cluster17}
\end{figure}

In Figure~\ref{fig:cluster17}(a), 'Information services' and 'Miscellaneous business services' are located at the upstream end.
'Medical and other health services' is also positioned upstream, while 'Social insurance, social welfare and care services' and 'Miscellaneous retail trade' appear downstream.
This arrangement suggests that information and business services support the operation and delivery of healthcare-related activities, with welfare and care and retail functions located closer to final service provision.
Figure~\ref{fig:cluster17}(b) shows that the cluster is concentrated in major metropolitan prefectures, particularly Tokyo and the surrounding Kanto area, as well as Osaka and other large urban regions.
This geographical pattern suggests that coordination-intensive information and business services are co-located with healthcare and related service delivery in urban areas.

\begin{figure}[htp]
    \centering
    \includegraphics[width=\linewidth]{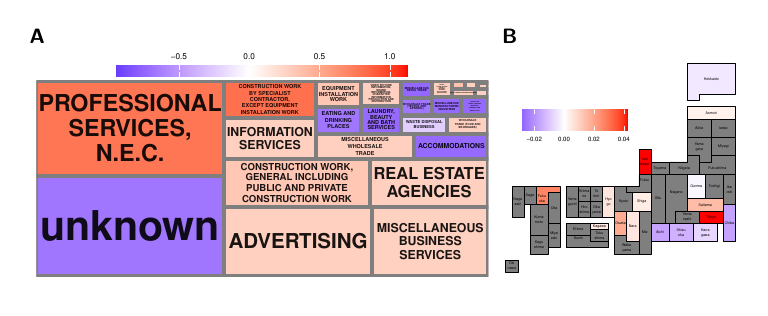}
    \caption{
    Industrial and geographic characteristics of Cluster 18.
    Panels \textbf{(A)}--\textbf{(B)} and the visual encodings are defined as in Fig.~1.
    }
    \label{fig:cluster18}
\end{figure}

In Figure~\ref{fig:cluster18}(a), 'Professional services, n.e.c.' and 'Construction work by specialist contractor, except equipment installation work' are positioned upstream, along with coordination- and market-facing sectors such as 'Advertising', 'Information services', 'Miscellaneous business services', 'Real estate agencies', and 'Miscellaneous wholesale trade'.
'Construction work, general including public and private construction work' and 'Equipment installation work' are placed in the upstream-to-intermediate range.
Downstream positions are occupied by consumer-facing sectors such as 'Accommodations', 'Eating and drinking places', and 'Laundry, beauty and bath services'.
Overall, the cluster is organized such that upstream professional, informational, and construction-related sectors are linked to the delivery and operation of urban facilities, while downstream hospitality and personal services correspond to end-use demand.
Figure~\ref{fig:cluster18}(b) indicates a geographical concentration in Tokyo, Osaka, and Fukuoka, suggesting that the coordination-intensive services and construction activities in this cluster are associated with major urban markets.
Ishikawa is an exception: although it is geographically distant from these metropolitan areas, it shows a notably high value.

\begin{figure}[htp]
    \centering
    \includegraphics[width=\linewidth]{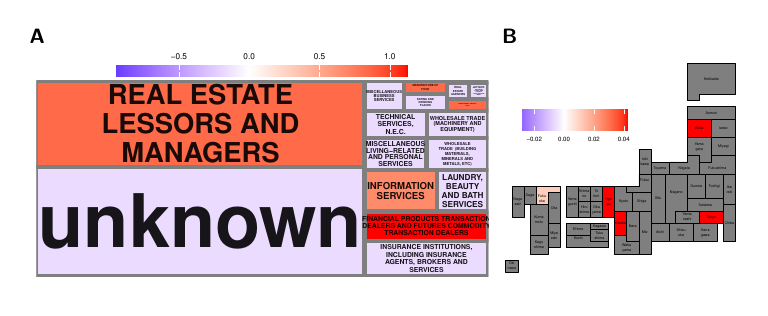}
    \caption{
    Industrial and geographic characteristics of Cluster 19.
    Panels \textbf{(A)}--\textbf{(B)} and the visual encodings are defined as in Fig.~1.
    }
    \label{fig:cluster19}
\end{figure}

In Figure~\ref{fig:cluster19}(a), 'Financial products transaction dealers and futures commodity transaction dealers' and 'Real estate lessors and managers' are placed at the upstream end, with 'Information services' also in a supporting position.
The downstream side includes 'Eating and drinking places', 'Laundry, beauty and bath services', and 'Miscellaneous living-related and personal services'.
'Insurance institutions, including insurance agents, brokers and services' and 'Real estate agencies' are also located on the consumer-facing side.
Figure~\ref{fig:cluster19}(b) shows a concentration in Tokyo, Osaka, Hyogo, and Fukuoka, all of which are major urban centers.
Akita is an exception: although it is geographically distant from these metropolitan areas, it shows a notably high value.

\begin{figure}[htp]
    \centering
    \includegraphics[width=\linewidth]{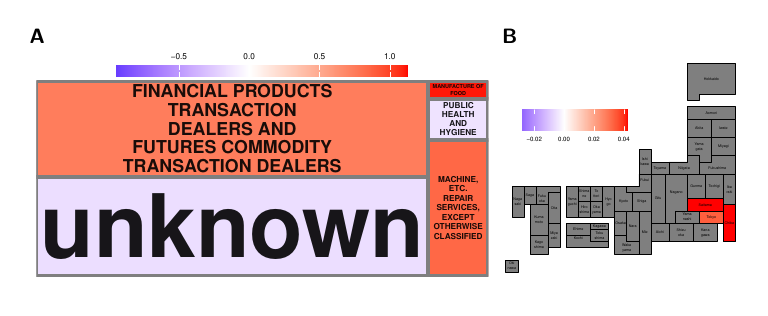}
    \caption{
    Industrial and geographic characteristics of Cluster 20.
    Panels \textbf{(A)}--\textbf{(B)} and the visual encodings are defined as in Fig.~1.
    }
    \label{fig:cluster20}
\end{figure}

As shown in Figure~\ref{fig:cluster20}(a), 'Financial products transaction dealers and futures commodity transaction dealers' has the largest share of total transaction volume and occupies an upstream position.
'Manufacture of food' is positioned even further upstream.
'Machine, etc. repair services, except otherwise classified' is also located in an upstream position.
'Public health and hygiene' is positioned downstream.
Because unclassified firms occupy a large share, the interpretation of this cluster is limited.
Figure~\ref{fig:cluster20}(b) shows that Saitama and Chiba have the highest relative densities, with Tokyo also showing an elevated value.
This indicates that the cluster is geographically confined to the capital region.

\begin{figure}[htp]
    \centering
    \includegraphics[width=\linewidth]{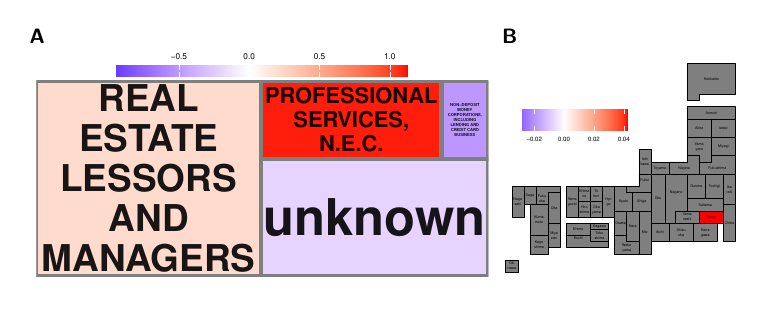}
    \caption{
    Industrial and geographic characteristics of Cluster 21.
    Panels \textbf{(A)}--\textbf{(B)} and the visual encodings are defined as in Fig.~1.
    }
    \label{fig:cluster21}
\end{figure}

Cluster 21 is composed of a small number of sectors, primarily real estate and professional services.
As shown in Figure~\ref{fig:cluster21}(a), 'Real estate lessors and managers' has the largest share and shows a near-neutral to slightly upstream position.
'Professional services, n.e.c.' has the strongest upstream position.
'Non-deposit money corporations, including lending and credit card business' is positioned on the downstream side.
Figure~\ref{fig:cluster21}(b) shows that this cluster is concentrated exclusively in Tokyo, and no firms are identified in other prefectures.
This pattern suggests that the cluster is associated with the concentration of real estate management and professional services in Tokyo.

\begin{figure}[htp]
    \centering
    \includegraphics[width=\linewidth]{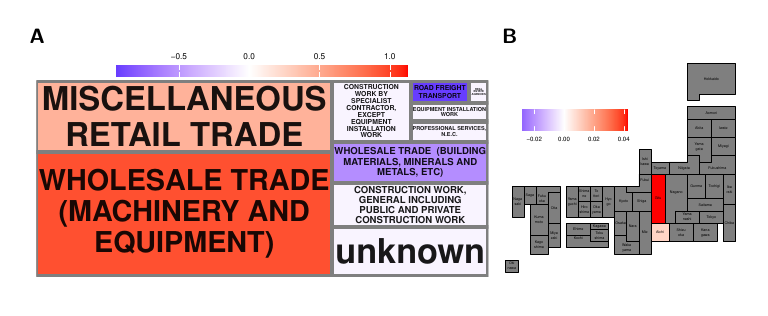}
    \caption{
    Industrial and geographic characteristics of Cluster 22.
    Panels \textbf{(A)}--\textbf{(B)} and the visual encodings are defined as in Fig.~1.
    }
    \label{fig:cluster22}
\end{figure}

As shown in Figure~\ref{fig:cluster22}(a), 'Wholesale trade (machinery and equipment)' has the largest share and occupies an upstream position.
'Miscellaneous retail trade' is also positioned upstream.
At intermediate positions, several construction-related sectors appear.
At the downstream end, 'Wholesale trade (building materials, minerals and metals, etc.)' and 'Road freight transport' occupy downstream positions.
This composition suggests that building material distributors and transport firms receive goods from machinery and equipment wholesalers, with construction-related activities connecting them.
Figure~\ref{fig:cluster22}(b) shows that the cluster is concentrated in only two prefectures: Gifu and Aichi.
This geographic distribution suggests a regionally specific construction and wholesale network in the Tokai area.

\begin{figure}[htp]
    \centering
    \includegraphics[width=\linewidth]{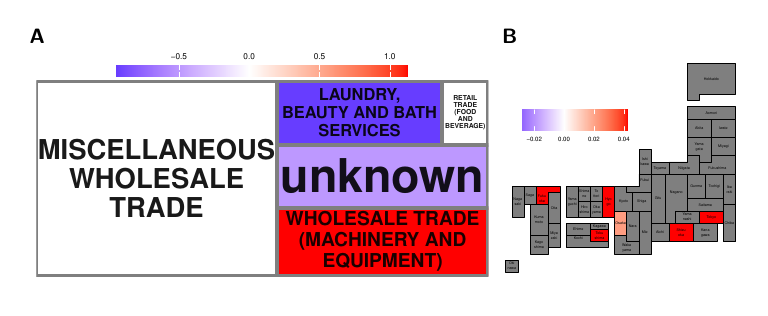}
    \caption{
    Industrial and geographic characteristics of Cluster 23.
    Panels \textbf{(A)}--\textbf{(B)} and the visual encodings are defined as in Fig.~1.
    }
    \label{fig:cluster23}
\end{figure}

In Figure~\ref{fig:cluster23}(a), on the upstream side, 'Wholesale trade (machinery and equipment)' is positioned furthest upstream, suggesting that machinery wholesalers function as suppliers to the other sectors.
'Miscellaneous wholesale trade' has the largest share and is located at an intermediate position, as is 'Retail trade (food and beverage)'.
'Laundry, beauty and bath services' is located on the downstream side.
Cluster~23 is similar to Clusters~14 and 16, in that it is related to personal-life services.
Figure~\ref{fig:cluster23}(b) shows that this cluster is present in several dispersed prefectures, including Fukuoka, Tokushima, Hyogo, Shizuoka, Tokyo, and Osaka.

\begin{figure}[htp]
    \centering
    \includegraphics[width=\linewidth]{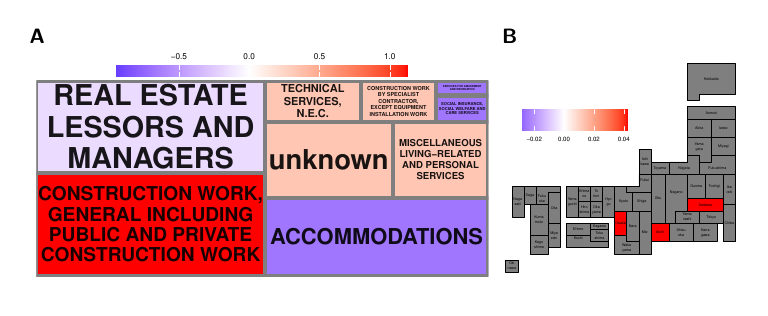}
    \caption{
    Industrial and geographic characteristics of Cluster 24.
    Panels \textbf{(A)}--\textbf{(B)} and the visual encodings are defined as in Fig.~1.
    }
    \label{fig:cluster24}
\end{figure}

'Construction work, general including public and private construction work' is positioned upstream in Figure~\ref{fig:cluster24}(a).
'Technical services, n.e.c.', 'Miscellaneous living-related and personal services,' and 'Construction work by specialist contractor' are at mid-to-upstream positions.
On the downstream side, 'Accommodations' has a strongly negative potential, along with 'Social insurance, social welfare and care services'.
This structure suggests that construction, technical, and living-related services supply the accommodation and welfare sectors.
As shown in Figure~\ref{fig:cluster24}(b), this cluster is concentrated in Saitama, Aichi, and Osaka, indicating a selective metropolitan distribution that does not include Tokyo.

\begin{figure}[htp]
    \centering
    \includegraphics[width=\linewidth]{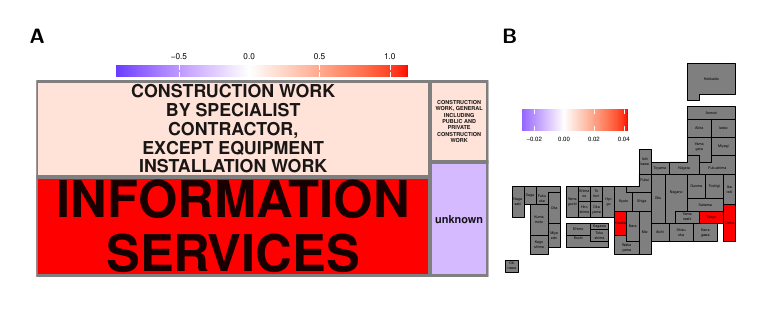}
    \caption{
    Industrial and geographic characteristics of Cluster 25.
    Panels \textbf{(A)}--\textbf{(B)} and the visual encodings are defined as in Fig.~1.
    }
    \label{fig:cluster25}
\end{figure}

As shown in Figure~\ref{fig:cluster25}(a), 'Information services' has the largest share and occupies an upstream position.
'Construction work by specialist contractor, except equipment installation work' and 'Construction work, general including public and private construction work' are at mid-to-upstream positions.
On the downstream side, firms with unidentified sectors, labelled as 'Unknown', show negative potentials.
Figure~\ref{fig:cluster25}(b) shows that this cluster is concentrated in Tokyo, Chiba, and Osaka.
Because the correspondence between upstream and downstream sectors is unclear, the interpretation of this cluster is limited.

\begin{figure}[htp]
    \centering
    \includegraphics[width=\linewidth]{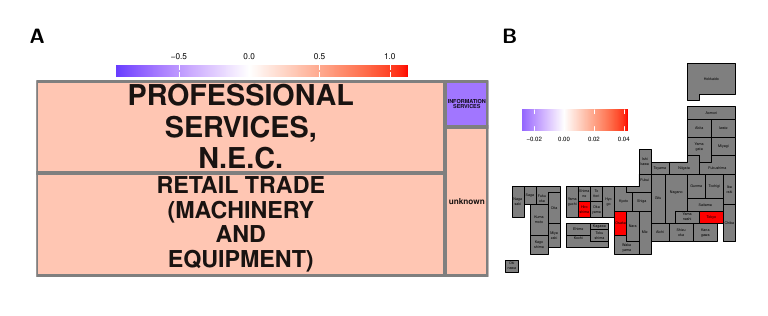}
    \caption{
    Industrial and geographic characteristics of Cluster 26.
    Panels \textbf{(A)}--\textbf{(B)} and the visual encodings are defined as in Fig.~1.
    }
    \label{fig:cluster26}
\end{figure}

Cluster 26 consists of a limited number of sectors, with professional services as the dominant component.
In Figure~\ref{fig:cluster26}(a), 'Professional services, n.e.c.' and 'Retail trade (machinery and equipment)' have the largest shares and occupy upstream positions.
'Information services' is the only sector in a downstream position.
Given the limited sectoral diversity, this cluster likely represents a small professional services group with minor retail and information service components.
Figure~\ref{fig:cluster26}(b) shows that this cluster is concentrated in Tokyo, Osaka, and Hiroshima.

\begin{figure}[htp]
    \centering
    \includegraphics[width=\linewidth]{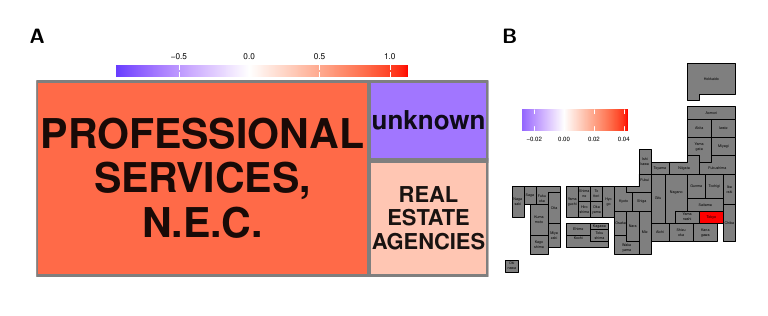}
    \caption{
    Industrial and geographic characteristics of Cluster 27.
    Panels \textbf{(A)}--\textbf{(B)} and the visual encodings are defined as in Fig.~1.
    }
    \label{fig:cluster27}
\end{figure}

Cluster 27 is the smallest cluster, consisting of only six firms.
As shown in Figure~\ref{fig:cluster27}(a), 'Professional services, n.e.c.' accounts for the majority of transaction volume and occupies an upstream position.
'Real estate agencies' is at a near-neutral to slightly positive position.
'Unknown' (firms with unidentified sectors) is positioned downstream.
Because the correspondence between upstream and downstream sectors is unclear, the interpretation of this cluster is limited.
Figure~\ref{fig:cluster27}(b) shows that all firms in this cluster are located in Tokyo, indicating that the cluster is specific to Tokyo.

\end{document}